# Superconducting nanowire single photon detectors for quantum information and communications

Zhen Wang, Shigehito Miki, and Mikio Fujiwara

*Abstract*— Superconducting nanowire single photon detectors (SNSPD or SSPD) are highly promising devices in the growing field of quantum information and communications technology. We have developed a practical SSPD system with our superconducting thin films and devices fabrication, optical coupling packaging, and cryogenic technology. The SSPD system consists of six-channel SSPD devices and a compact Gifford-McMahon (GM) cryocooler, and can operate continuously on 100 V ac power without the need for any cryogens. The SSPD devices were fabricated from high-quality niobium nitride (NbN) ultra-thin films that were epitaxially grown on single-crystal MgO substrates. The packaged SSPD devices were temperature stabilized to 2.96 K +/- 10 mK. The system detection efficiency for an SSPD device with an area of 20 × 20 μm$^2$ was found to be 2.6% and 4.5% at wavelengths of 1550 and 1310 nm, respectively, at a dark count rate of 100 c/s, and a jitter of 100 ps full width at half maximum (FWHM). We also performed ultra-fast BB84 quantum key distribution (QKD) field testing and entanglement-based QKD experiments using these SSPD devices.

## I. Introduction

SINGLE photon detectors are one of the key elements in the field of quantum information and communications technology. For these applications, the photon detectors should have high speed, high quantum efficiency, low dark count rate, and low timing jitter. The detector performance has a direct impact, for example, on the distance, the speed and the security level of quantum key distribution (QKD). At telecommunication wavelengths, InGaAs/InP avalanche photodiodes (APDs) have been widely used for detection of single photons [1]-[3]. However, they exhibit high dark counts, and must be operated at gated Geiger mode with long dead time to avoid after pulse phenomena. This would often complicate a total communications system design.

Superconducting photon detectors are highly promising competitors to conventional semiconductor APDs. They do not require any gated operation, which makes it useful for constructing a compact and stable system for quantum communications and quantum information experiments. Two types of superconducting photon detectors have been developed and extensively applied to quantum information and communications applications. One is the transition edge sensor (TES), and the other is superconducting nanowire single photon detector (SNSPD or often simply SSPD). TESs have very low dark count rates and high detection efficiencies (DE) of 80% for wide wavelengths [4], and even exceed 90% at 1550 nm wavelength [5] at a low operating temperature of 100 mK. These characteristics are useful to improve the performance of QKD as demonstrated in [6]over a distance of 67.5 km. The TESs used there had a timing resolution of 90 ns full width at half maximum (FWHM) and a dead time of 4 μs. The other superconducting detector, SSPD, has higher timing resolution and shorter dead time, and hence would be useful for high clock rate systems although the DE remains lower and needs to be improved [7]-[10]. SSPDs are typically made of NbN nanowire, and The operating temperature is around the liquid He temperature of 4.2 K. Presently, SSPD systems with a compact cryocooler have been developed and successfully employed in QKD experiments for extending the transmission distance as well as the key generation rate [11]-[15]. The DE was typically around a few % at 1550 nm wavelength and the dark count rate was about 100 c/s [16],[17]. Dauler et al. have recently achieved the detection efficiency as high as 50% at 1550 nm wavelength and the dark count rate of 1 kc/s [18],[19].

Typical current performances of semiconductor based single photon detectors and SSPDs are summarized in TABLE I. As a figure of merit, we may use the performance index defined by (DE)/((Dark count rate)(Jitter)). In terms of this measure, SSPDs are promising even with relatively small detection efficiency, because they have very low dark counts and low jitter.

However, further improvements in device performance are highly desirable and will broaden the impact of SSPDs in QKD and other quantum information processing applications. In particular, significant effort is being put into increasing the detection efficiency and photon counting rate. In order to

Manuscript received February 5, 2009.

Zhen Wang is with the Kobe Advanced Research Center, National Institute of Information and Communications Technology, 588-2, Iwaoka, Nishi-ku, Kobe, Hyogo 651-2492, Japan (corresponding author to provide e-mail: wang@nict.go.jp).

Shigehito Miki is with the Kobe Advanced Research Center, National Institute of Information and Communications Technology, 588-2, Iwaoka, Nishi-ku, Kobe, Hyogo 651-2492, Japan (e-mail: s-miki@nict.go.jp).

Mikio Fujiwara is with the National Institute of Information and Communications Technology, 4-2-1 Nukui-Kitamachi, Koganei, Tokyo 184-8795, Japan (e-mail: fujiwara@nict.go.jp).

TABLE I
PERFORMANCES OF VARIOUS SINGLE PHOTON DETECTORS AT 1500 nm WAVELENGTH

| Detectors | DE (%) | Dark count (c/s) | After pulse probability | Count rate (Hz) | Jitter (ps) | Performance index (×10⁻⁶) | Operation mode |
|---|---|---|---|---|---|---|---|
| InGaAs/InP APD [1] (Sinusoidally gate) | 5.1 | 7.6 K | 2.3% | 20M | 100 | 6.7 | Gated (1.5 GHz) |
| InGaAs/InP APD [2] (Self-differencer) | 10.8 | 2.9 K | 6.16% | 100M | 55 | 6.8 | Gated (1.25 GHz) |
| SFG Si APD [3] | 6 | 10 K | not reported | 100M | 75 | 8 | Continuous (PPLN up conversion) |
| SSPD[a] | 2 | 30 | 0 | 66 M | 100 | 660 | Continuous |

[a]The performances depend on operation conditions, device design, and so on. The data is from typical values with our 10 × 10 μm² device.

achieve these goals, the technical issues that need to be improved are deposition of superconducting ultra-thin films with high quality and uniformity, fabrication of nanowires with good homogeneity, and optical coupling between SSPD devices and fibers with high efficiency. Furthermore, the development of a multichannel SSPD system with a compact, turnkey, cryogen-free cooler is essential for practical application.

We have systematically developed and investigated a niobium nitride (NbN) superconducting single photon detectors system with Gifford-McMahon (GM) cryocooler, and have employed it in QKD experiments. In this paper, we describe in detail the fabrication of high-quality NbN thin films and SSPD devices, measurement of kinetic inductance (KI), optical packaging technique, development and testing of the SSPS system with the GM cryocooler, and results of QKD experiments with the SSPD devices.

## II. FABRICATION OF SSPD DEVICES

### A. Deposition and properties of NbN thin films

So far, NbN has been chosen for SSPD devices due to its relatively high transition temperature, and ease of fabricating ultrathin films and nanostructure. The electric properties of NbN thin films, however, are very strongly dependent on the crystal structure, leading to difficulties in reducing the KI and controlling device parameters. One possible way of solving the problem is to use single-crystal NbN thin films [20]. Therefore, we started this work by fabricating single-crystal NbN thin films for stabilizing film quality.

The NbN thin films were deposited by reactive dc-magnetron sputtering in a mixture of Ar and $N_2$ gases at ambient temperature. The background pressure was below $1.5 \times 10^{-7}$ Torr, and the total pressure was set at 2 mTorr to elevate the sputtering energy. The relative amounts of Ar and $N_2$ introduced for sputtering were carefully controlled to the ratio 5:1 by mass flow controllers. The target was 99.99 % pure niobium and the target size was a diameter of 8 in. Single-crystal MgO (100) substrates with a thickness of 0.4 mm were used to promote the epitaxial growth of the films. A dc power supply was used to stabilize the discharge state, and the bias current was set to 3.0 A. A detailed explanation of the deposition process and the method used to find the optimum bias conditions may be found elsewhere [21],[22].

Figure 1 plots $T_c$ and the resistivity at 20 K ($\rho_{20K}$) as a function of film thickness for our NbN thin films. Though a fall in $T_c$ and increase in $\rho_{20K}$ occurred with a reduction in film thickness, excellent superconducting properties with a $T_c$ of 9.7 K, residual resistivity ratio (= $\rho_{300K}/\rho_{20K}$) of 0.87 and $\rho_{20K}$ of 185 μΩcm were obtained even in a 2.8 nm thick film. Figure 2 shows a cross-sectional transmission electron micrograph (TEM) of NbN films on MgO substrates. The deposited NbN films were epitaxially grown as single crystal with their (200) plane set parallel to the substrate. It is evident that the single-crystal NbN films grew directly on the MgO substrates without any initial amorphous layers being formed. The lattice matrix was found to be well ordered, and no columnar structures were observed in the region observed by TEM.

### B. Device fabrication and dc characteristics

The nanowire meanders of the SSPD devices were fabricated by direct e-beam lithography and reactive ion etching (RIE). Optimized e-beam lithography conditions made it possible to fabricate meander lines with widths of 80-100 nm. The filling factor, which is the ratio between the line and space, was always set to 50% in this work. Figure 3 presents a NbN nanowire with a width of 100 nm and a meander line area of 20 × 20 μm² observed through (a) microphotograph, (b) scanning electron micrograph (SEM), and (c) TEM. Coplanar waveguide (CPW) lines with an input impedance of 50 Ω were connected to the nanowire to read the output signal. These were fabricated by standard photolithography and a lift-off process. Since NbN ultrathin films break easily from damage during fabrication and thermal stress near the electrodes [23], 150 nm thick NbN thin films were used for the CPW lines. These introduce minimal stress on the NbN ultrathin film meander lines and have relatively strong adherence. In addition, the surface resistance of NbN thin films is as low as that of any metal at several gigahertz and is sufficient to transmit the output signal.

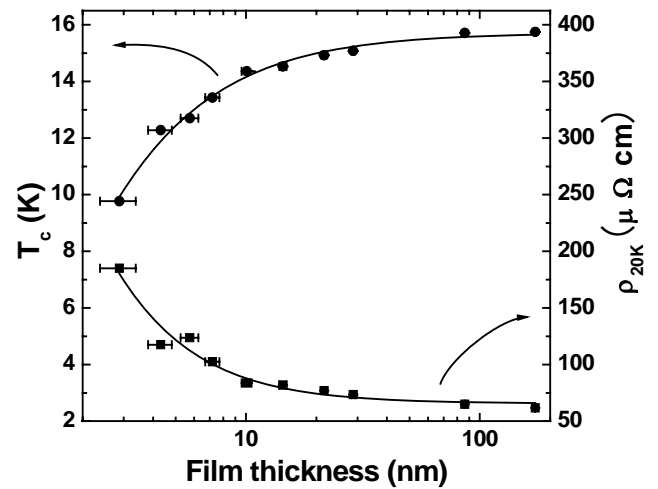

Fig. 1. $T_c$ and $\rho_{20K}$ of NbN thin films as a function of films thickness

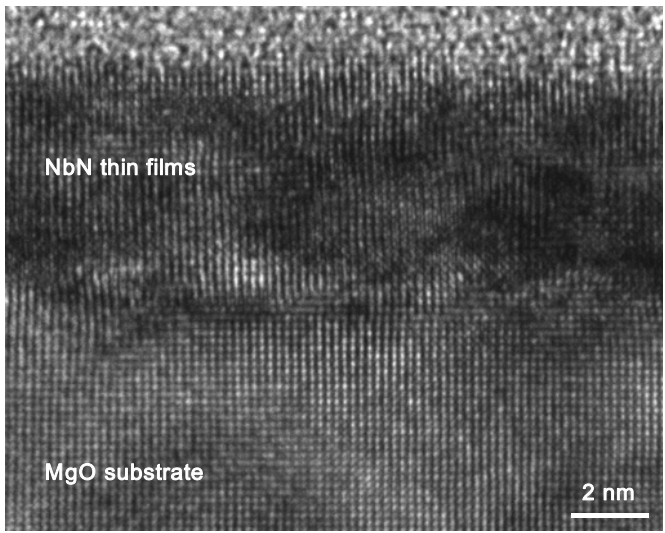

Fig. 2. Cross-sectional transmission electron micrograph (TEM) of NbN thin films deposited on MgO substrates. The thickness of NbN thin films was 5 nm.

The dc characteristics of the SSPD devices were measured at 4.2 K. Figure 4(a) and (b) show the I-V and R-T characteristics of a typical NbN SSPD device, respectively. A clear critical current and sharp transition can be seen in the figure. Figure 4 (c) shows histograms of the measured sheet resistance at 20 K ($R_{\text{sheet,20K}}$) and critical current density ($J_c$), of our 80 and 100 nm wide meanders in a film grown under identical conditions on an MgO substrate. A total of five (seven) samples of 80- (100-) nm-wide meanders were measured. $R_{\text{sheet,20K}}$ and $J_c$ were determined from the measured resistance at 20 K ($R_{20K}$) and the critical current ($I_c$) of the meanders by using the design geometry. As shown in Fig. 4(b), $R_{\text{sheet,20K}}$ was almost constant regardless of the sample and design, indicating that the meanders were fabricated as designed in a reproducible manner. However, the $J_c$ of the devices, which will be limited by local defects anywhere in the nanowire, were smaller for the 80 nm wide meanders than the 100 nm wide meanders. This discrepancy in $J_c$ indicates that the 80 nm wide meanders have more constrictions than the wider meanders; this is expected because the narrow meander is 1.25 times longer. However, the variation among like-width meanders remained within ±5%, implying a high reproducibility of patterning.

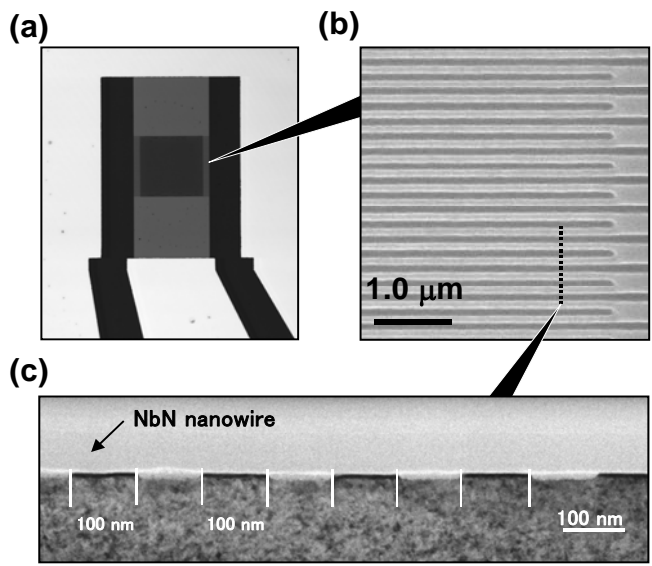

Fig. 3. (a) Micro photograph of -, (b) Scanning electron micrograph (SEM) of -, and (c) Transmission electron micrograph (TEM) of SSPD device.

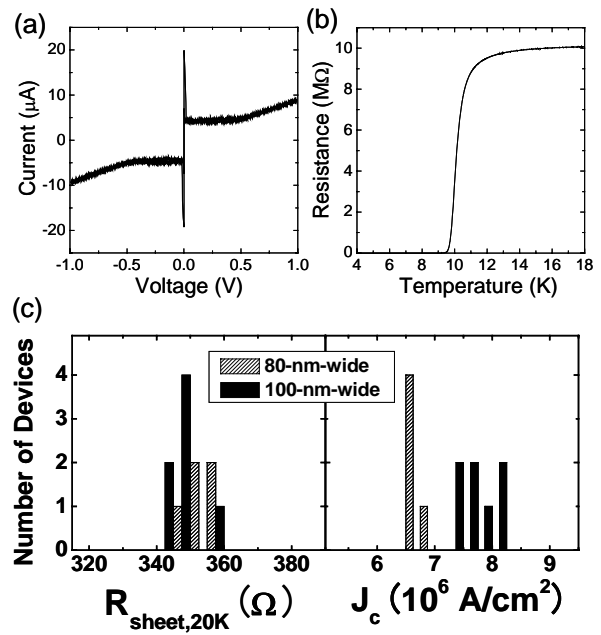

Fig. 4. DC properties of NbN-SSPDs. (a) I-V curve of -, and (b) R-T curve of our typical SSPD device. (c) Histograms of sheet resistance and $J_c$ of our 5 (7) SSPDs with 80 (100) nm wide meanders of 4.2 nm thick NbN films on a common MgO substrate.

### III. DEVELOPMENT OF SSPD SYSTEM

#### A. System setup

For practical applications, the development of a multichannel SSPD system with a compact, turnkey, cryogen-free cooler is essential. Over the past decade, commercially available closed-cycle cryocoolers have improved both in terms of attainable base temperature and reliability. The smallest available unit provides 0.1 W of cooling at 4 K. This is sufficient to cool multiple detector packages. Several versions of this system have now been constructed and are in use in Japan, USA, and Europe

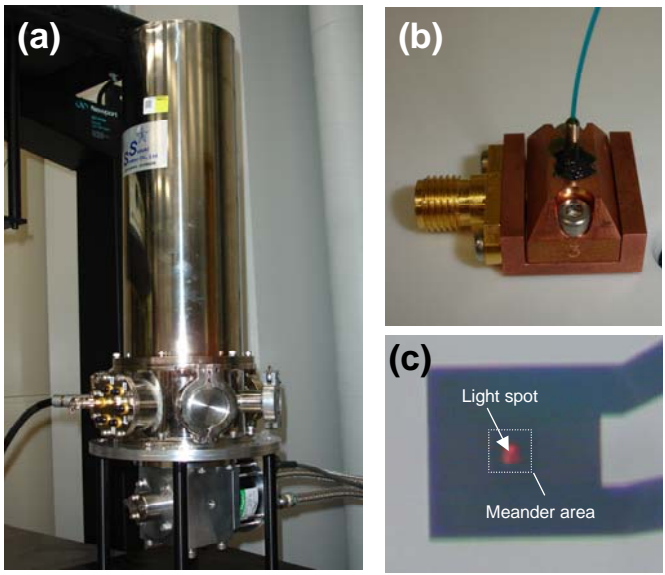

Fig. 5. Photographs of (a) SSPD system with 3 K GM cryo-cooler system, (b) SSPD packages for fiber coupling, and (c) illuminated spot and SSPD meander area after careful alignment and packaging

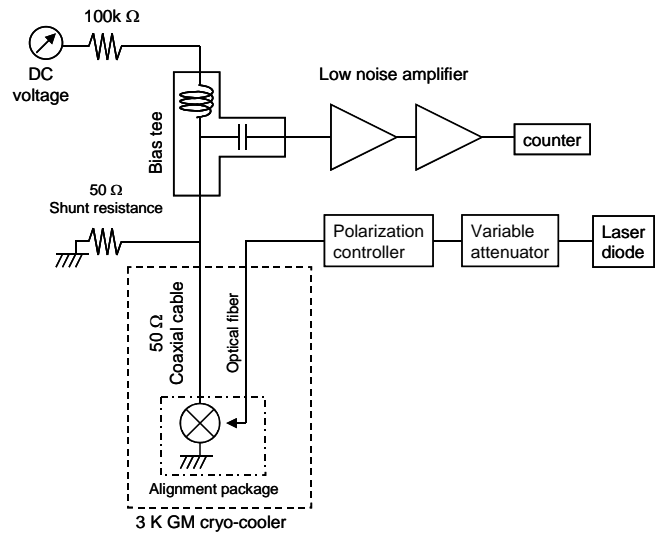

Fig. 6. Schematic view of setup with 3 K GM cryo-cooler based SSPD system for single photon detection measurement.

[16],[17],[24]. In our SSPD system, a two-stage small GM cryocooler was used. Figure 5(a) presents a photograph of a cooler stage on which six SSPD packages (shown in Fig. 5(b)) can be mounted. The sample stage used for cooling the SSPD packages was connected to the second stage through a stainless steel plate and a lead block with a large heat capacity to reduce thermal fluctuation [16]. It could be cooled to 2.96 K with a thermal fluctuation range of 10 mK in a short time range. To achieve efficient coupling between the incident photons and the meander area, we prepared SSPD packages consisting of a pair of oxygen-free copper blocks: one for mounting the device and the other to fix the end of an single mode optical fiber with a core diameter of 9 μm. Prior to cooling, the blocks were fixed so that the end of the fiber is accurately aligned with the center of the meander area. This was performed by monitoring the device and light spot from the backside of the substrate, as shown in Fig. 5(c). The distance between the end of fiber and meander area was kept between 20–80 μm. After careful adjustment, the SSPD packages were set on the sample stage.

Figure 6 shows the measurement setup used for testing the performance of our SSPD system. Each SSPD package was current-biased via the dc arm of a bias tee, and the output signal was counted through the ac arm of the bias tee and two low noise amplifiers. Single photons that acted as inputs were introduced by single-mode optical fibers. A 1550 (1310) nm wavelength continuous wave laser diode was used as the input photon source; it was heavily attenuated so that the photon flux at the input connector of the cryostat was $10^7$ photons/s, which was accurately calibrated by optical power meter. A polarization controller was inserted in front of the optical input port to control the polarization properties of the incident photons so that their polarization sensitivity (maximizing the DE) matched that of each device. The output port from the nanowire was connected to a parallel shunt resistance (50 Ω) and the bias tee at room temperature through a 50 Ω co-axial cable. The shunt resistance was used to prevent the biased device from latching. The system DE was defined as the output count rate divided by the photon flux rate input to the system.

### B. System performance

The nanowire geometry, dc characteristics, and device performance measured by our SSPD system for two (plus one in Ref. [25]) typical device designs are listed in TABLE II. Since the thickness of the nanowire was less than that described in Ref. [25], the devices showed lower $T_c$ and $I_c$. Figure 7 plots the system DE and dark count rate as a function of the bias current normalized by $I_c$ for an SSPD device with design #A. The system DE (including coupling loss) at a dark count rate of 100 c/s was about 2.6% and 4.5% at 1550 and 1310 nm, respectively. These values were almost the same as our previous results [25], taking into account the higher operation temperature. The DE for an ideal SSPD device is dependent principally on the absorption and geometrical filling factor of nanowire devices. However, it is necessary to consider various factors sequentially to improve the system DE for real devices, such as device design (nanowire width, size, thickness, and filling factor), extent of constrictions [26], and operating temperature. For example, a nanowire design not optimized for higher sensitivity and constriction in a nanowire would limit the system DE since there is no saturation region in the bias dependencies of the system DE, as shown in Fig. 7. Further optimization of nanowire design and reduction in constrictions will facilitate improvements in device DE.

TABLE II
NANOWIRE DESIGN, NANOWIRE DC CHARACTERISTICS, AND MEASURED DEVICE PERFORMANCE FOR TWO (PLUS ONE REFERRED IN [25]) TYPICAL SSPD DESIGNS.

|  | Ref [25] | #A | #B |
| --- | --- | --- | --- |
| Device size (μm) | 20 × 20 | 20 × 20 | 10 × 10 |
| Line width (nm) | 100 | 100 | 100 |
| Line pitch (nm) | 200 | 200 | 200 |
| films thickness (nm) | 4.2 | 3.9 | 3.9 |
| $T_c$ (K) | 12 | 9.8 | 9.5 |
| $I_c$ (μA) | 32 | 19 | 15 |
| $R_{20K}$ (MΩ) | 7 | ~10 | ~2.5 |
| System DE (%) | 0.4-3.5 at 2.5K | 0.8-2.6 at 2.95 K | 1.0 - 2.6 at 2.95 K |
| $L_{k,device}$ (μH) | 1.0 | 1.3 | 0.3 |

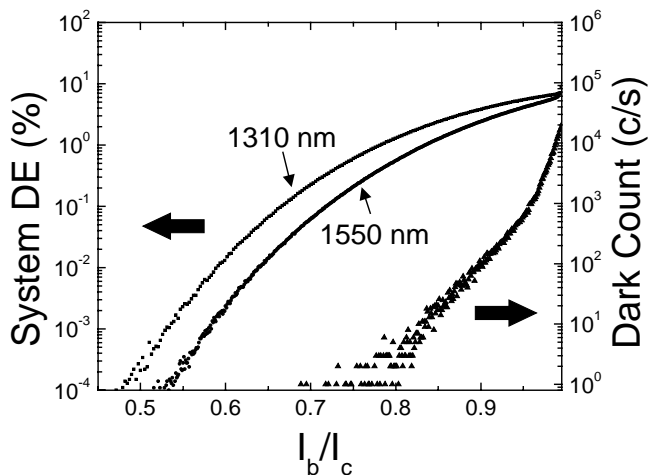

Fig. 7. System DE and dark count rate as a function of bias current normalized by $I_c$ for a SSPD device with design #A (20 × 20 μm² area device).

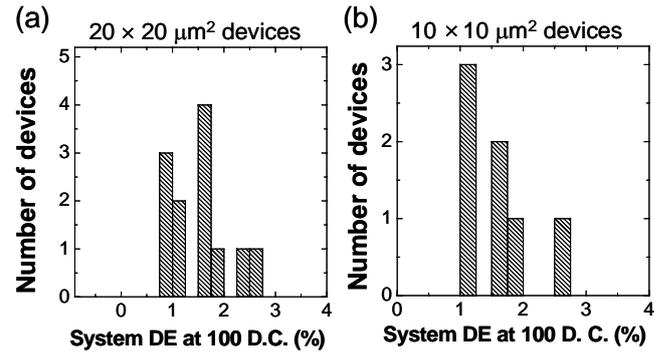

Fig. 8. Histogram of system DE at 100 c/s dark count rate for (a) 12 measured devices with design #A (20 × 20 μm² area device) and (b) 7 measured devices with design #B (10 × 10 μm² area device).

For practical applications, a SSPD system with multiple detectors is essential. Figure 8 shows a histogram of the system DE at 100 c/s dark count rate for 12 devices with 20 × 20 μm² active area (Fig. 8(a)) and 7 devices with 10 × 10 μm² active area (Fig. 8(b)). Fiber coupling efficiencies were estimated to be almost 100 % because the distances between the end of fibers and devices were kept within 50 μm (20 μm) for 20 × 20 μm² (10 × 10 μm²) area devices. System DEs exceeding 1% were obtained in all devices, as shown in Fig. 8, implying high stability and repeatability of our fabricating process. The 10 × 10 μm² devices, in spite of small area, showed similar performances to 20 × 20 μm² devices with a maximum system DE of 2.5 %. In addition, the kinetic inductance of the devices $L_{k,device}$, measured by observing phase of a reflected microwave signal versus frequency using a network analyzer [25], can be reduced from 2.0 μH to 0.3 μH with downsizing of device area. This brings practical advantage in increasing operating speed of the SSPDs for practical applications[10],[27].

We also investigated the stability of our SSPD system by running it for over 10 h (from 12:00 to 22:00). Figure 9 plots the system DE, dark count, and device temperature as a function of the running time. The SSPD device was biased at a dc current to achieve a dark count rate of 100 c/s. As shown in the figure, the SSPD system was able to operate with a constant system DE of 2.6% and dark count rate of 100 c/s within the measuring time. A sudden increase in the dark count was observed on a rare occasion. We think that it may be result from disturbance from the external environment or ground noise during the day. The temperature fluctuation range within the measuring time was about 50 mK which was greater than 10 mK described in section III. It is considered that the variation of system power supply voltage caused this additional fluctuation.

A timing jitter of a 10 × 10 μm² device was measured by using a time-correlated single photon counting card in a desktop PC operated on the histogram mode. The SSPD pulse starts a timer and a trigger pulse from a pulse laser with 33 MHz repetition rate stops. Illuminated photon pulses were heavy attenuated and average photon numbers was less than one photon. Figure 10 is the profile of the jitter measurement. As shown in Fig. 10, our SSPD device with 10 × 10 μm² detection area offers low timing jitter of 100 ps FWHM at 1550 nm. In the application of QKD, such detectors with significant low jitter enable to reduce quantum bit error rate from pulse dispersion in the transmission medium. Therefore, our SSPDs are promising candidates for QKD at high (gigahertz) clock rates.

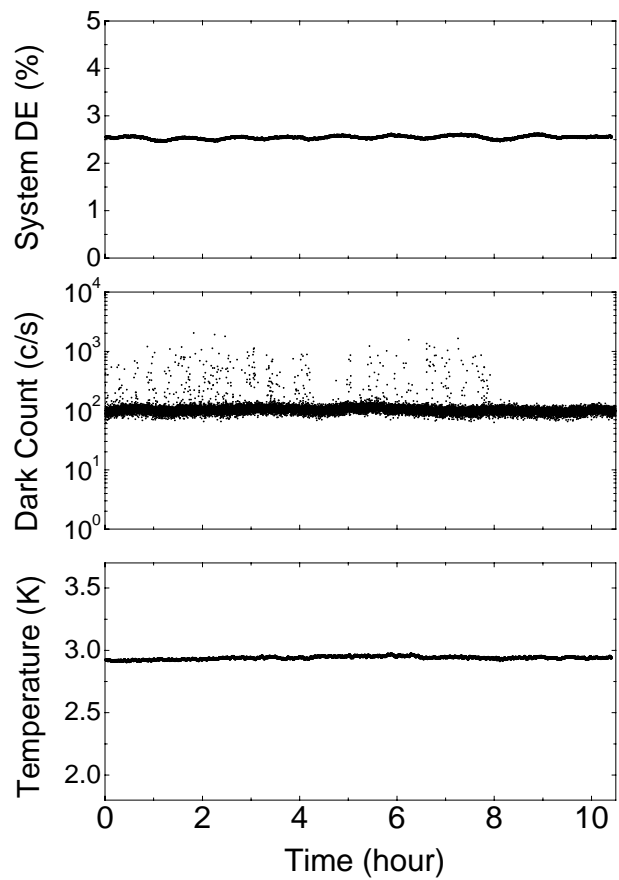

Fig. 9. System DE, dark count rate, and operation temperature of SSPD system as a function of the running time. Device with design #A (20 × 20 μm² area device) was mounted into the system.

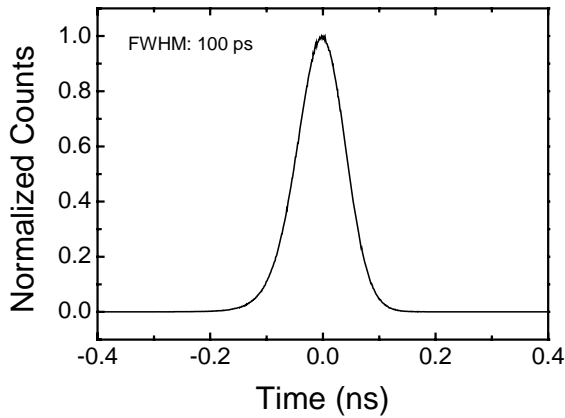

Fig. 10. Timing jitter measurements of a 10 × 10 μm² SSPD device illuminated with 1550 nm wavelength.

## IV. QKD EXPERIMENTS

SSPDs that provide high speed single photon detection for the 1550 nm telecom-band described above have been successfully employed in QKD experiments. Here we briefly mention two experiments. One is a field experiment of ultra fast BB84-QKD transmission through a 97 km installed fiber [13]. The SSPD system with four NbN nanowires made by Moscow State Pedagogical University and National Institute of Information and Communications Technology (NICT), and packaged by National Institute of Standards and Technology (NIST), was used to receive the quantum signal. The detection efficiency and the dark count rate of these SSPDs were 1.2-1.6% and 90-160 count/s, respectively. This system was combined with a one way BB84-QKD system developed by NEC, which is based on planar light circuit interferometers, and worked at 625 MHz clock rate using practical clock synchronization based on wavelength-division multiplexing (WDM). We succeeded in over-one-hour stable key generation at a high sifted key rate of 2.4 kbps and a low quantum bit error rate (QBER) of 2.9%. The asymptotic secure key rate was estimated to be 0.78-0.82 kbps from the transmission data with the decoy method of average photon numbers 0, 0.15, and 0.4 photons/pulse.

The other is an experiment of BBM92 QKD with time-bin entangled photon pairs over a 100 km optical fiber in laboratory [14]. The same SSPD system was combined with an efficient entangled photon pair source that consists of a fiber coupled periodically poled lithium niobate waveguide and ultra low loss filters, and planar light circuit interferometers. The experiment lasted approximately 8 hours stably, and successfully generated a 16 kbit sifted key with a quantum bit error rate of 6.9 % at a rate of 0.59 bits per second, from which we were able to distill a 3.9 kbit secure key.

## V. CONCLUSION

We have successfully developed a superconducting single photon detectors system, and have applied it to QKD experiments. In spite of the fact that the performance currently achieved by our SSPD devices and system are far short of their intrinsic potential, the total system performance (DE, dark count rate, speed, and jitter) is still better than that of commercial APD units at the telecommunication wavelengths. Their high performance also proved extremely useful in the QKD experiments. We must explore many avenues to realize the high potential of the SSPD devices and system, for example, by increasing the DE using an integrated optical cavity and anti-reflection coating, reducing the KI by downsizing the active area and making array elements. We anticipate high performance in the near future in both SSPD devices and practical systems. We believe that they will be widely used not only in the fields of quantum information and quantum communications but also in the fields of optical[28], particle, and molecular sensing.

## ACKNOWLEDGMENT

We would like to thank Sae Woo Nam, Burm Baek at National Institute of Standards and Technology, Aaron J. Miller at Albion College, and Robert H. Hadfield at Heriot-Watt University for helpful advice, discussions, and collaboration work to this study.

We also would like to thank Akihiro Tanaka, Yoshihiro Nambu, Seigo Takahashi, Wakako Maeda, Ken-ichiro Yoshino, Akio Tajima, and Akihisa Tomita at NEC corporation, Toshimori Honjo, Hiroki Takesue, Qiang Zhang, Hidehiko Kamada, Yoshiki Nishida, Osamu Tadanaga, and Masaki Asobe at NTT corporation, Kyo Inoue at Osaka university and Yoshihisa Yamamoto at Stanford university for fruitful collaboration work about BB84 and BBM92 QKD experiment in section VI, respectively.

We would like to acknowledge Masahide Sasaki through whole work in this paper (development of devices, system, QKD experiments).